\documentclass[prb,twocolumn,floatfix,showpacs,amsmath,amssymb]{revtex4}
\usepackage{graphicx}
\usepackage{graphics}
\usepackage{pstricks}
\usepackage[latin1]{inputenc}

\newcommand{\cmas}{c^{\dagger}}
\newcommand{\cmin}{c}

\newcommand{\spin}{\vec{S}}

\begin{document}

\title{ Comparison between disordered quantum spin $1/2$ chains}
\author{ C.A.\ Lamas }
\affiliation{Departamento de F\'{\i}sica, Universidad Nacional de La Plata, C.C.\ 67, 1900 La
Plata, Argentina.}

\author{ D.C.\ Cabra }
\affiliation{Departamento de F\'{\i}sica, Universidad Nacional de La Plata, C.C.\ 67, 1900 La
Plata, Argentina.}
\affiliation{Laboratoire de Physique Th\'{e}orique, Universit\'{e} Louis
Pasteur, 3 Rue de l'Universit\'{e}, 67084 Strasbourg, C\'edex, France.}
\affiliation{Facultad de
Ingenier\'\i a, Universidad Nacional de Lomas de Zamora, Cno.\ de Cintura y Juan XXIII, (1832)
Lomas de Zamora, Argentina.}

\author{ M.D.\ Grynberg }
\affiliation{Departamento de F\'{\i}sica, Universidad Nacional de La Plata, C.C.\ 67, 1900 La
Plata, Argentina.}

\author{ G.L.\ Rossini }
\affiliation{Departamento de F\'{\i}sica, Universidad Nacional de La Plata, C.C.\ 67, 1900 La
Plata, Argentina.}

%PHYSICAL REVIEW B \hspace{1.4cm}\textit{VOLUME xx, NUMBER xx}\hspace{2.2cm} \today

%
\begin{abstract}
We study the magnetic properties of two types of one dimensional
$XX$ spin $1/2$ chains. The first type has only nearest neighbor
interactions which can be either antiferromagnetic or
ferromagnetic and the second type which has both nearest neighbor
and next nearest neighbor interactions, but only antiferromagnetic
in character. We study these systems in the presence of low
transverse magnetic fields both analytically and numerically.
Comparison of results show a close relation between the two
systems, which is in agreement with results previously found in
Heisenberg chains by means of a numerical real space
renormalization group procedure.
\end{abstract}

% 75.10.Jm Quantized spin models
% 75.10.Nr Spin-glass and other random models
% 75.40.Mg Numerical simulation studies
\maketitle

\section{INTRODUCTION}
\label{intro}

One-dimensional quantum spin systems have been extensively studied over the last years
\cite{Giamarchi}. In particular, randomness has a profound effect on their physical properties
and is always present in real systems through impurities or structural disorder. This can even
produce singular behaviors in the magnetic properties, not observed in pure systems. One of the
main motivations to study disordered quantum spin chains is the possibility of classifying their
behavior in universality classes associated to different regions in their phase diagrams \cite{Hoyos,Yusuf}.

In the last few years, numerical works \cite{Hoyos,Yusuf} have
shown that the thermodynamic properties of disordered Heisenberg
chains with nearest neighbors (NN) and next nearest neighbors
(NNN) couplings, {\it both antiferromagnetic}, are very similar to
those found in disordered chains with {\it only} NN couplings
which can be {\it either antiferromagnetic or ferromagnetic}. In fact,
it was shown that under Real Space Renormalization Group (RSRG)
the former systems flow to a fixed point characterized as a chain
with only NN couplings in a given distribution, taking both antiferromagnetic and ferromagnetic
values.

More specifically, let us consider a Heisenberg chain with Hamiltonian
\begin{eqnarray}
H=\sum_{i}\left(J_{i}\spin_{i} \cdot \spin_{i+1}+
J'_{i}\spin_{i} \cdot \spin_{i+2}\right),
\end{eqnarray}
where $\spin_{i}$ are spin-$1/2$ operators and the couplings
$J_{i}>0$ and $J'_{i}>0$, both antiferromagnetic, follow
probability distributions $P(J_{i})$, $P(J'_{i})$.
\begin{figure}[hpt]
\begin{pspicture}*(0.7,0.8)(8.5,3.6)
%\psgrid
\psline[linecolor=gray]{-}(2,2)(8,2)
%atomitos
\pscircle*[linecolor=black](2,2){0.1}
\pscircle*[linecolor=black](3,2){0.1}
\pscircle*[linecolor=black](4,2){0.1}
\pscircle*[linecolor=black](5,2){0.1}
\pscircle*[linecolor=black](6,2){0.1}
\pscircle*[linecolor=black](7,2){0.1}
\pscircle*[linecolor=black](8,2){0.1}
%los operadores S
\rput(2,1.7){\footnotesize $\spin_1$}
\rput(3,1.7){\footnotesize $\spin_2$}
\rput(4,1.7){\footnotesize $\spin_3$}
\rput(5,1.7){\footnotesize $\spin_4$}
\rput(6,1.7){\footnotesize $\spin_5$}
\rput(7,1.7){\footnotesize $\spin_6$}
\rput(8,1.7){\footnotesize $\spin_7$}
%
%the stronger coupling
\psline[linecolor=black,linewidth=2pt]{-}(4.1,2)(4.9,2)
%algunos couplings
\rput(4.5,2.2){\footnotesize $J_{34}$}
%\rput(5.5,2.4){$J_2$}
%\rput(6.5,2.4){$J_3$}
%
%other interactions
\pscurve[linecolor=black,linestyle=dotted,linewidth=1.5pt]{-}(2,2.1)(2.5,2.4)(3,2.53)(3.5,2.4)(4,2.1)
\pscurve[linecolor=black,linestyle=dotted,linewidth=1.5pt]{-}(4,2.1)(4.5,2.4)(5,2.53)(5.5,2.4)(6,2.1)
\pscurve[linecolor=black,linestyle=dotted,linewidth=1.5pt]{-}(6,2.1)(6.5,2.4)(7,2.53)(7.5,2.4)(8,2.1)
\pscurve[linecolor=black,linestyle=dotted,linewidth=1.5pt]{-}(3,1.9)(3.5,1.6)(4,1.47)(4.5,1.6)(5,1.9)
\pscurve[linecolor=black,linestyle=dotted,linewidth=1.5pt]{-}(5,1.9)(5.5,1.6)(6,1.47)(6.5,1.6)(7,1.9)
%\rput(9,2.2){$J>0$}
%
\end{pspicture}
\label{chain} \caption{Schematic picture of a
disordered-antiferromagnetic NN-NNN chain.}
\end{figure}
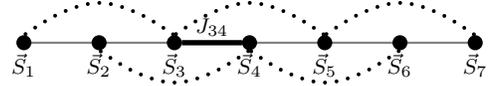
Let us review the arguments of Ref.\ [\onlinecite{Hoyos,Yusuf}]. If we
consider the adjacent spins that are coupled by the strongest bond
(say the spins 3 and 4 in the figure \ref{chain}), and its
neighbors, we have a problem whose Hamiltonian can be written as
\begin{eqnarray}
    H=H_{0}+H_{I}+H_{rest},
\end{eqnarray}
where
\begin{eqnarray}
    H_{0}&=&J_{34}\spin_{3}\cdot \spin_{4},
\end{eqnarray}
\begin{eqnarray}
    \nonumber H_{I}&=& J_{23}\spin_{2}\cdot \spin_{3}  +  J_{45}\spin_{4}\cdot \spin_{5}
      +   J'_{35}\spin_{3}\cdot \spin_{5}\\
    &+&     J'_{24}\spin_{2}\cdot \spin_{4} .
\end{eqnarray}
and $H_{rest}$ corresponds to all the other spins that are not
coupled to spins 3 and 4.

The sole consideration of  $H_{0}+H_{I}$ is enough to determine the
effective interaction between the spins 2 and 5, as follows
\cite{Dasgupta,Fisher,Yusuf}. The ground state for the Hamiltonian
$H_{0}$ is a singlet with energy $E_0=-\frac{3}{4}J_{34}$ while
excited triplet states have energy $E_1 = \frac{1}{4}J_{34}$.
Since $J_{34}$ is the largest bond, one can take $H_{I}$ as a
perturbation. Regarding $\spin_{2}$ and $\spin_{5}$ as external
operators, a second-order perturbation calculation gives an
effective Hamiltonian describing the low energy sector (we
consider only the coupling generated between 2 and 5)
\begin{eqnarray}
    \nonumber E&=&-\frac{3}{4}J_{34}-\frac{3}{16J_{34}}[(J_{23}-J'_{24})^{2}+(J'_{35}-J_{45})^{2}]\\
    &+&\frac{(J_{23}-J'_{24})(J_{45}-J'_{35})}{2J_{34}}\spin_{2}\cdot \spin_{5}.
\end{eqnarray}
From this result one can remove the spins $\spin_{3}$, $\spin_{4}$
from the original Hamiltonian, replacing them by an effective NN
coupling $\tilde{J}_{25}$
\begin{eqnarray}
    \tilde{J}_{25}=\frac{(J_{23}-J'_{24})(J_{45}-J'_{35})}{2J_{34}},
\end{eqnarray}
between $\spin_{2}$ and $\spin_{5}$.

Following this RSRG decimation procedure one ends up with a NN
spin $1/2$ chain in which effective interactions like
$\tilde{J}_{25}$ can be ferromagnetic. Notice that when the NNN
couplings are very weak compared to the NN couplings the
ferromagnetic effective interaction is unlikely to appear. On the
contrary, for strong NNN couplings, RSRG does generate effective
ferromagnetic couplings and the system flows to a phase
controlled by large effective spins at low energies \cite{Yusuf}.

It is the aim of the present note to further understand the
connection between these and related systems, in more general
situations. In particular we are interested in comparing magnetic
properties of easy plane $XX$ spin $1/2$ chains in the presence of
uniform magnetic fields.

We first study analytically the magnetic properties of quantum
$XX$ spin $1/2$ chains with antiferromagnetic and ferromagnetic NN
interactions, using an elegant argument relating magnetization
with random walk problems \cite{egg}. In a second step we
implement a numerical self consistent method based in a mean field
approximation \cite{rossini}, which allows for analyzing quantum
$XX$ spin $1/2$ chains with NN and NNN  interactions, to be
applied in the antiferromagnetic case. Finally, common features
found in these spin systems are discussed.

The structure of the paper is the following. Analytical results
derived from the random walk argument are presented in section
\ref{af-F} for homogeneous and dimerized disorder in
antiferromagnetic NN chains, as well as novel results for
dimerized distributions of antiferromagnetic and ferromagnetic
couplings in NN chains. The self consistent numerical method is
presented in Section \ref{scmf}, then tested on pure (ordered) NNN
antiferromagnetic chains in Section \ref{pure}, and finally applied
to the NNN disordered case in Section \ref{disorder}. Numerical
results are compared with exact diagonalization for small chains,
up to 24 spins, at each step.
%A numerical study of  spin-spin
%correlation functions is presented in Section \ref{corr}.
The comparison between both types of chains, summary and
conclusions are presented in Section \ref{summary}.
%
%
%
%
%%%%%%%%%%%%%%%%%%%%%%%%%%%%%%%%%%%%%%%%%%%%%%%%%%%%%%%%%%%%%%%%%%%%%%%%%%%%%%%%%%%%%%%%%%%%%%%%%%%%%%%%%%%%%%%%%%%%%%%% AF-F %%%%%%%%%%%%%%%%%%%%%%%%%%%%%%%%%%%%%%%%%%%%%%%%%%%%%%%%
%
\section{Analytical treatment for NN spin $1/2$ chains}
\label{af-F}
In this Section we study analytically the magnetization of one dimensional
spin $1/2$ systems with NN interactions whose Hamiltonian in the $XX$ model is
\begin{eqnarray}\label{espin xx}
    H=\sum_{i}J_{i}(S^{x}_{i}S^{x}_{i+1}+ S^{y}_{i}S^{y}_{i+1})-h\sum_{i}S_{i}^{z} ,
\end{eqnarray}
where $J_{i}$ are random nearest neighbor couplings (either antiferromagnetic (AF) or
ferromagnetic (F)) and $h$ is a uniform magnetic field. By means of the
Jordan-Wigner transformation \cite{JW}
\begin{eqnarray}\label{j-w}
        \nonumber S^+_i & \equiv & c^\dagger_i e^{i\pi \phi_i},\\
        S^-_i & \equiv  & e^{-i\pi \phi_i}c_i ,\\
        \nonumber S^{z}_i & \equiv & c^\dagger_i c_i - \frac{1}{2},
\end{eqnarray}
where $\phi_i  \equiv  \sum_{l=1}^{i-1}\cmas_l \cmin_l $,
we can rewrite this Hamiltonian in terms of spinless fermionic operators
\begin{eqnarray}
\label{fer_1D}
H=\sum_i t_i \left( c^\dagger_i c_{i+1} + {\rm
H.c.} \right)-h\sum_{i}c^\dagger_i c_i ,
\end{eqnarray}
(in this Section we use $t=J/2$). Notice that the magnetic field acts
as a chemical potential for the fermions.

We start with the study of the homogeneous disordered case where
the couplings follow a homogeneous distribution
($P(t_{i})=P(t_{j})\ \forall i,j$). Following the argument
presented in Ref.\ [\onlinecite{egg}] one can introduce a random
variable $u_i$ which, for a {\em one} particle eigenstate of the
Hamiltonian in eq.\ (\ref{fer_1D}), undergoes a random walk
behavior between a reflecting barrier at
$u_{max}=\log(\tilde{t}^2/E)$ and  an absorbing barrier at
$u_{min}=\log(E)$ with $E$ being the energy of the eigenstate and
$\tilde{t}$ the positive average value of AF couplings $t_i$. The
relevant quantity to compute here is the number of eigenstates
$\textbf{\emph{N}}(E)$ with energies below $E$, which for low
energies is related to the average number $\bar{n}$ of steps
necessary to complete a diffusion cycle from the reflecting
barrier to the absorbing one. The relation between them is
\cite{schmidt}
\begin{eqnarray}
\textbf{\emph{N}}(E)=\frac{1}{2\bar{n}}+\frac{1}{2}.
\end{eqnarray}
In the present case one gets $\bar{n}\sim(u_{max}-u_{min})^{2}/\sigma^{2}$,
where $\sigma^2$ is the variance of the coupling distribution $P$,
rendering
\begin{eqnarray}\label{num_de_est2}
    \textbf{\emph{N}}(E)\cong \frac{1}{2}\left( 1+\frac{\sigma^{2}}{(\ln(\tilde{t}/E)^{2})^{2}} \right).
\end{eqnarray}
This result indeed describes the magnetization of the spin system (\ref{espin xx}).
We write the mean magnetization $m$ in terms of fermionic variables using the Jordan-Wigner transformation
\begin{eqnarray}
    m=\langle \sum_{i}S^z_i\rangle=\sum_{i}(\langle c^\dagger_{i}, c_{i}\rangle-1/2),
\end{eqnarray}
in order to exhibit the relation between magnetization and fermion filling.
Recalling that the magnetic field acts as a chemical potential and regulates the fermion filling,
for low magnetic field we finally obtain a mean magnetization per site
\begin{eqnarray}
\label{num_de_est}
M(h)\sim \frac{1}{2}\left( \frac{\sigma^{2}}{[ \ln(\tilde{t}/h)^{2}]^{2}} \right)
\end{eqnarray}
(here $M=2m/N$, with $N$ the number of spins in the chain, is normalized to $1$ at saturation).
%% dimerized AF %%

Now we turn to the case of dimerized inhomogeneous distributions
($(P_{odd}(t_i)$ for odd sites $i$ different from  $P_{even}(t_j)$ for even sites $j$).
The result above was suitably generalized in Ref.\ [\onlinecite{Cabra}] for this case:
the random variable $u_{i}$ undergoes a random walk with diffusion coefficient
$D$ and a driving force $F$ given by
\begin{eqnarray}
    \label{d}
    D&=&\frac{1}{2}[ \text{var}_{odd}^{2}(\log(t^{2}_{i}))+2\text{var}_{even}^{2}(\log(t^{2}_{j}))],
    \\
    \label{f}
    F&=&\langle \log(t_i^{2})\rangle_{odd}-\langle \log(t_j^{2})\rangle_{even},
\end{eqnarray}
where $\text{var}$ stands for the variance of the corresponding distribution.
The average numbers of steps for a diffusion cycle to
be completed  behaves now as $\bar{n}\sim e^{\alpha \ (u_{max}-u_{min})/2}$.
Then the magnetization of the system is seen to follow a power law
\begin{eqnarray}
    M(h)\sim h^{\alpha},
\end{eqnarray}
with $\alpha=\frac{2F}{D}$.

%% F-AF %%

In what follows we generalize this procedure to study the system
of interest here, namely a disordered spin $1/2$ chain with
AF and F NN interactions. Let us
consider the following binary coupling distribution
\begin{eqnarray}
\label{d_com}
        P(t_i)= xP_{F}(t_i)+(1-x)P_{AF}(t_i)\,,
\end{eqnarray}
with weight $x$ for F couplings and $1-x$ for AF ones, combined
with dimerization in the sense described above (both $P_{F}$ and
$P_{AF}$ are different for even and odd sites). A similar pattern
for disorder was proposed in recent numerical studies
\cite{Hoyos,Yusuf}.

We can again map the system onto a random walk problem with
driving force and appropriate barriers. In particular the driving
force can be written in terms of the single distribution
parameters as
%
%\scriptsize
\begin{eqnarray}
        \nonumber F = x \langle  \log( (t_{i_{odd}})^2/(t_{i_{even}})^2) \rangle_{F}+
        \nonumber \\
        (1-x) \langle \log ( (t_{i_{odd}})^2(t_{i_{even}})^2) \rangle_{AF},
\end{eqnarray}
where subindexes $odd$ and $even$ indicate the
distribution to be used for disorder average.

Notice that even under the strong hypothesis that both  single
distributions are dimerized (inhomogeneous)
$${\small \langle  \log(  (t_{i_{odd}})^2/(t_{i_{even}})^2) \rangle_{F}\neq 0},$$
and
$${\small\langle \log ( (t_{i_{odd}})^2(t_{i_{even}})^2) \rangle_{AF}\neq 0},$$
the competition between AF and F couplings can eventually cancel the driving force,
under the condition
\begin{eqnarray}
   \lefteqn{  \langle \log(t^{2}_{i_{odd}}/t^{2}_{i_{even}})\rangle_{AF}=}  \\
   & & \nonumber
   x[\langle \log(t^{2}_{i_{odd}}/t^{2}_{i_{even}})\rangle_{AF}
    -\langle \log(t^{2}_{i_{odd}}/t^{2}_{i_{even}})\rangle_{F}].
    \label{condition}
\end{eqnarray}
This shows that there are two phases present in the system.
For the coupling distribution in eq.\ (\ref{d_com}) we have that,
at least for low magnetic field $h$, the magnetization follows a power law
$M\sim h^{\alpha}$ in most of the parameter space,
while there exists a line, with $x$ satisfying eq.\ (\ref{condition}),
where the magnetization is logarithmic in $h$, $M\sim \frac{1}{\log(h^{2})^{2}}$.

It is important to stress that in the power law regime the dynamical exponent $\alpha$
can be larger or smaller than one.
For the present case it is still given by $2F/D$ where $F$ and $D$ are computed as in
eqs.\ (\ref{f}) and (\ref{d}) but with the binary even and odd distributions in eq.\ (\ref{d_com}).
When the disorder parameter (in this case var$(\ln(t^{2}))$) is small
and the dimerization is no longer considered, the dynamical exponent
turns out to be larger than one.
In contrast, when the disorder is
stronger the dynamical  exponents take values smaller than one and
the magnetic susceptibility displays a {\em singularity} at the origin.
It is worth to point out that this behavior was also reported
by RSRG studies in dimerized NN chains \cite{Hyman}.

%%%%%%%%%%%%%%%%%%%%%%%%%%%%%%%%%%%%%%%%%%%%%%%%%%%%%%%%%%%%%%%%%%%%%%%%%%%%%%%%%%%%%%%%%%%%
\section{Numerical self consistent treatment for antiferromagnetic NNN spin $1/2$ chains}
\label{scmf}
%%%%%%%%%%%%%%%%%%%%%%%%%%%%%%%%%%%%%%%%%%%%%%%%%%%%%%%%%%%%%%%%%%%%%%%%%%%%%%%%%%%%%%%%%%%%

In this Section we consider a spin 1/2 $XX$ Hamiltonian in $d=1$ with both NN and NNN
AF position dependent interactions under a uniform magnetic field, namely
\begin{eqnarray}
 \nonumber
H&=&\sum_{i=1}^{N} [J_{i}(S_{i}^{x}S_{i+1}^{x}+S_{i}^{y}S_{i+1}^{y})\\
&+& J'_{i}(S_{i}^{x}S_{i+2}^{x}+S_{i}^{y}S_{i+2}^{y}) ]
 - h \sum_{i=1}^N S^z_{i} \label{xxz} \,,
\end{eqnarray}
where $N$ is the number of spins in the chain and periodic
boundary conditions are assumed. The inclusion of NNN couplings does not allow
the analytical procedure used above.
We then perform a numerical self consistent mean field (SCMF) study of this system.

We first review the procedure proposed in \cite{rossini} and then
apply it to our present case.
In terms of the fermion operators in eq.\ (\ref{j-w}) this Hamiltonian reads
%
%\footnotesize
\begin{eqnarray}
\label{fer2}
\nonumber
H &=& \sum_{i=1}^{N}\frac{J_{i}}{2}\big[c^\dagger_{i}c_{i+1}+{\rm H.c.} \big]+\\
&+&\sum_{i=1}^{N}\frac{J'_{i}}{2}
\big[e^{-i\pi \hat{n}_{i+1}}c^\dagger_{i}c_{i+2}+ {\rm H.c.} \big]+\\
\nonumber &-&h\sum_{i=1}^{N}(c^\dagger_{i}c_{i}-\frac{1}{2}) \,.
\end{eqnarray}
%\normalsize
%
First, in order to enable a
single particle treatment of $H$, we approximate the local
fermionic occupation numbers $\hat n_{i}=\cmas_{i}\cmin_{i}$ by
their expectation values in an arbitrarily chosen initial state to
be varied and determined self consistently. The local parameters
$n_{i}$ satisfy the constraint $\sum_{i=1}^{N}(n_{i}-1/2)=m$, with
$m$ the system magnetization. Then, in this mean field (MF)
approximation the Hamiltonian can be written as a quadratic form
\begin{eqnarray}
\label{cuadratico}
    H_{XX}^{(MF)}(\{n_i\})=\sum_{i,j} c^\dagger_i J_{ij}(\{n_i\})c_j ,
\end{eqnarray}
where
%\footnotesize
\[
J_{ij}(\{n_{i}\})=\left\{ \begin{array}{lcl}
                        \displaystyle{\frac{J_{i}}{2}} & \mbox{if} & i,j \ \mbox{are  NN,} \\
                            &                       &           \\
                        \displaystyle{\frac{J'_{i}}{2}\ e^{i\pi n_{i+1}}} & \mbox{if} &  i,j \ \mbox{are NNN,} \\
                        & & \\
                        0 &  & \mbox{otherwise.} \\
                        \end{array}
        \right.
\]
\\
%\normalsize
We have omitted the Zeeman term
$h\sum_{i=1}^{N}(c^\dagger_{i}c_{i}-\frac{1}{2})$ as, being
diagonal, it can be added later on.

It is our aim to find an approximation to the actual ground state
(GS) at a given magnetization $m$. Thus, we estimate this state in
the MF Hamiltonian (\ref{cuadratico}) by solving the one particle
spectrum and filling the lowest energy-levels to satisfy the
constraint $\sum_{i=1}^{N}(n_{i}-1/2)=m$. Then, we compute a new
set of local parameters $n'_{i}=\langle
GS|\cmas_{i}\cmin_{i}|GS\rangle$ which we use again as input for
Eq. (\ref{cuadratico}). Iterating this procedure, we finally
obtain a fixed point configuration of occupation numbers
$n'_{i}(\{n_{p}\})=n_{p}$.

Specifically, the quadratic Hamiltonian can be written in  diagonal form
\begin{eqnarray}
    H=\sum_{k=1}^{N}\epsilon(k)d^\dagger_{k}d_{k} \,,
\end{eqnarray}
where the operators $c_{i}$ are related with $d_{k}$ by the
unitary transformation
\begin{eqnarray}
\label{unitary1} c_{i}=\sum_{k}d_{k}\bigl(Q^{\dagger}\bigr)_{ki},
\end{eqnarray}
where $Q_{ik}$ is the matrix of eigenvectors of
$J_{ij}(\{n_{i}\})$. Using standard methods \cite{NR} we can
easily compute the eigenvalues $\epsilon(k)$ of
$H_{XX}^{(MF)}(\{n_i\})$ and eigenvectors $Q_{ik}$ for fairly
large spin systems.
The set of $d_k$ satisfy fermion anticommutation relations $\{d_k
, d^\dagger_{k'}\}=\delta_{k,k'}$ and the total fermion number is
conserved
\begin{eqnarray}
    N_{f}=\sum_{i=1}^{N}c^\dagger_{i}c_{i}=\sum_{k=1}^{N}d^\dagger_{k}d_{k} \,.
\end{eqnarray}
So, the ground state with magnetization $m$, in the diagonal basis is given by
\begin{eqnarray}
\label{gs}
    |GS\rangle=\prod_{k=1}^{m+N/2}d^\dagger_{k}|0\rangle \,.
\end{eqnarray}
%%%
In this state the energy at zero temperature (for zero magnetic field) is simply given by

\begin{eqnarray}
    E_{GS}(m,0)=\sum_{k=1}^{m+N/2}\epsilon(k)\,,
\end{eqnarray}
whereas for $h \neq 0 $ the total energy is shifted as
$E_{GS}(m,h)=E_{GS}(m,0) - h \,m$. The magnetization curve $m(h)$ can finally be obtained
by minimizing the energy $E_{GS}(m,h)$ for different given magnetizations.
%

%%%%%%%%%%%%%%%%%%%%%%%%%%%%%%%%%%%%%%%%%%%%%%%%%%%%%%%%%%%%%%%%%%%%%%%%%%%%%%%%%%%%%%%%%%%%%%%%%%%%%%%%%%%%%%%%%%%%% PURECASE %%%%%%%%%%%%%%%%%%%%%%%%%%%%%%%%%%%%%%%%%%%%%%%%%%%%%%

\section{Results for the pure case}
\label{pure}
%%%%%%%%%%%%%%%%%%%%%%%%%%%%%%%%%%%%%%%%%%%%%%%%%%%%%%%%%%%%%%%%%%%%%%%%%%%%%%%%%%%%%%%%%%%%

\begin{figure}[t]
\begin{pspicture}*(-0.5,2.5)(12,8.5)
%\psgrid
\rput(3.8,5.5){\includegraphics[width=0.50\textwidth]{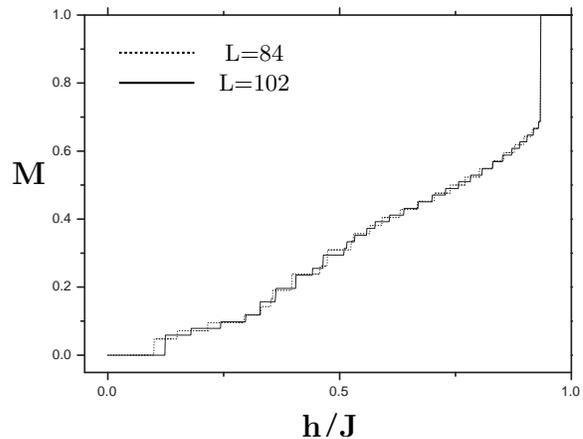}}
%
%\rput(5,5.8){L=12}
\rput(3,7.6){L=84}
\rput(3,7.2){L=102}
%\psline[linestyle=dotted]{-}(3.2,5.8)(4.2,5.8)
\psline[linestyle=dashed, dash=1pt 1pt]{-}(1.2,7.6)(2.2,7.6)
\psline{-}(1.2,7.2)(2.2,7.2)
%tapo el label que ten\'{\i}a
\pscustom[linestyle=none,fillstyle=solid,fillcolor=white]{
\psline(-1,2.6)(0.2,2.6)
\psline(0.2,2.6)(0.2,6)
\psline(0.2,6)(-1,6)
\psline(-1,6)(-1,2.6)
}
%le pongo un nuevo label
\rput(0,6){\large \textbf{M}}
\pscustom[linestyle=none,fillstyle=solid,fillcolor=white]{
\psline(-3,2.5)(5,2.5)
\psline(5,2.5)(5,3)
\psline(5,3)(3,3)
\psline(3,3)(3,2.5)
}
%le pongo un nuevo label
\rput(4,2.6){\large \textbf{h/J}}
%
%\rput(5,-0.2){Figura 4.2:\emph{ J'/J=0.6 se observa el plateau de magnetizaci\'on en M=0}}
%\rput(5,-1){\emph{ como el que vimos en el cap\'itulo \ref{sec:egg}.}}
\end{pspicture}
\caption{\label{plateau_scmf} Magnetization curves for pure chains with $L=102$ and $L=84$ sites
both with $J'/J=0.6$. For $M=0$ there is a magnetic plateau.}
\end{figure}

We have tested the SCMF procedure in a pure (ordered) $J-J'$ $XX$
model using chain lengths of up to the order of 100 spins, and compared its
results  with those found by exact diagonalization in smaller chains. In Fig.
\ref{plateau_scmf}, we show the magnetization behavior under a
magnetic field for $J'/J=0.6$ which clearly indicates the
existence of a magnetization plateau at $M=0$.

This initial plateau shows up only in a narrow region of $J'/J$ which
in the SCMF approximation was estimated within the bounds  $0.55 \alt J'/J \alt 0.75$.
No subsequent plateaus were observed in the system.
Notice that in other models (e.g the $XXZ$ model\cite{japones})
there is a plateau at $M=1/3$, however this not the case in the $XX$ situation.

\begin{figure}[t]
\begin{pspicture}*(1,0.5)(11,7)
%\psgrid
\centering
\rput(5,4){\includegraphics[width=0.40\textwidth]{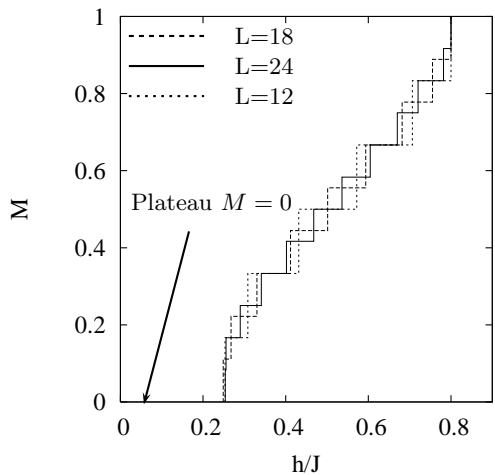}}
%\rput(5.5,0.0){Figura 4.6:\emph{ Curva de magnetizaci\'{o}n obtenida mediante   }}
%\rput(5,-0.4){\emph{Lanczos para $J_2/J_1=0.6$}}
\psline{<-}(3.4,1.7)(4,4)
\rput(4.3,4.4){Plateau $M=0$}
\rput(5,5.8){L=12}
\rput(5,6.6){L=18}
\rput(5,6.2){L=24}
\psline[linestyle=dashed,dash=1pt 2pt](3.2,5.8)(4.2,5.8)
\psline[linestyle=dashed,dash=2pt 2pt]{-}(3.2,6.6)(4.2,6.6)
\psline{-}(3.2,6.2)(4.2,6.2)
%\pscircle(3.4,3.2){0.5}
%
\end{pspicture}\\
\caption{\label{lanczos} Magnetization curve for small pure chains with $J'/J=0.6$,
obtained by exact diagonalization. There is a clear plateau at $M=0$.}
\end{figure}

\begin{figure}[t]
\begin{pspicture}*(1,0.5)(11,7)
%\psgrid
\centering
\rput(4.6,4){\includegraphics[width=0.40\textwidth]{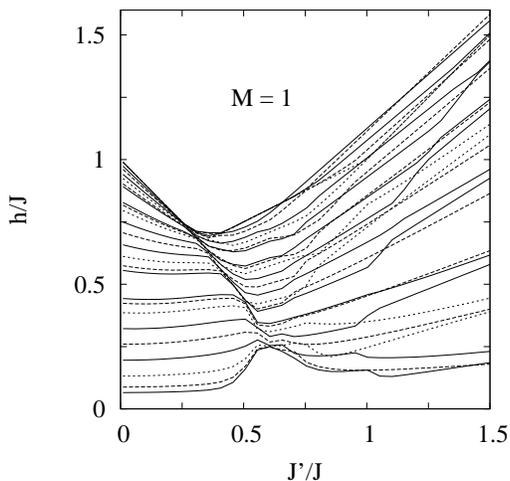}}
\end{pspicture}\\
\caption{\label{diag_lanczos} Magnetic phase diagram for pure chains. In the regime $0.5 \alt J'/J  \alt 0.75$
the critical field $h_c$ increases, showing the formation of the $M=0$ plateau. }
\end{figure}

To lend further support to our SCFM approach, we  compared the above results with those obtained
in smaller systems using exact diagonalization \cite{Golub}.
In Fig.\ \ref{lanczos} we exhibit the magnetization
curve obtained by  Lanczos technique using chains of 12, 18 and 24 spins,
where an $M=0$ plateau also emerges at $J'/J=0.6$. The regime where this plateau appears
turns out to be slightly higher than that found with SCMF.
This can be observed in Fig.\ \ref{diag_lanczos}
where we show the magnetic phase diagram obtained in a wide region of coupling parameters.
Here, each line stands for a critical field above which  the magnetization is increased by flipping one spin.
In agreement with our SCMF expectations, we can see that the first critical field
(lowest line of each studied length)
is higher in the region $0.5 \alt  J'/J \alt 0.75$ where the $M=0$ plateau is favored and size effects become
substantially reduced.

%
%%%%%%%%%%%%%%%%%%%%%%%%%%%%%%%%%%%%%%%%%%%%%%%%%%%%%%%%%%%%%%%%%%%%%%%%%%%%%%%%%%%%%%%%%%%%
\section{Results for the disordered case}
\label{disorder}
%%%%%%%%%%%%%%%%%%%%%%%%%%%%%%%%%%%%%%%%%%%%%%%%%%%%%%%%%%%%%%%%%%%%%%%%%%%%%%%%%%%%%%%%%%%%

We now apply the SCMF procedure to our main interest,
namely disordered antiferromagnetic $XX$ chains with a Hamiltonian given by eq.\ (\ref{xxz}).
In this case $J_{i}$ and $J'_{i}$ are random NN and NNN coupling exchanges respectively.
For concreteness, let us consider Gaussian distributions of exchanges
$ \displaystyle{P(J_{i})\propto e^{\frac{-(J_i-\bar J)^{2}}{2\sigma^{2}}}}$
with mean value $\bar J=J>0$ for NN couplings, $\bar J=J'>0$ for NNN couplings
and the same disorder strength $\sigma$ in both cases
(we explicitly forbid negative couplings in the Gaussian tail).
We compute numerically
the chain magnetization by averaging over many disorder realizations.
Typically we considered over 200 samples with periodic boundary conditions
and random sets of initial fermionic distributions.

\begin{figure}[t]
\begin{pspicture}*(1,1.2)(10,7)
% figura con j'/j=0.5 , sigma=0.5, desorden=100
%\psgrid
\rput(5,4){\includegraphics[width=0.45\textwidth]{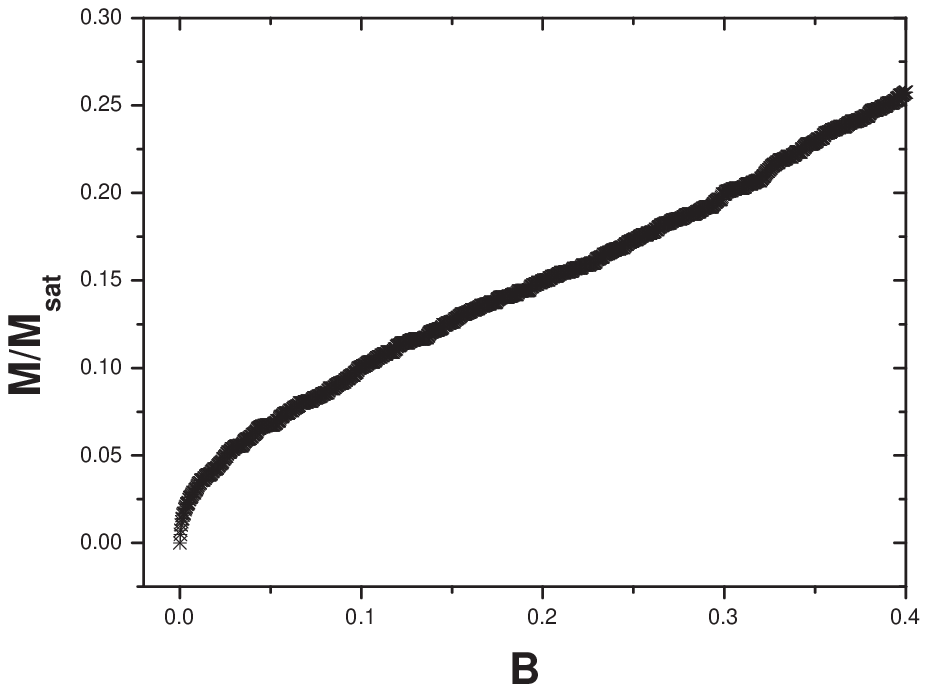}}
%tapo el label que ten\'{\i}a
\pscustom[linestyle=none,fillstyle=solid,fillcolor=white]{
\psline(0.6,2.6)(1.8,2.6)
\psline(1.8,2.6)(1.8,6)
\psline(1.8,6)(0.6,6)
\psline(0.6,6)(0.6,2.6)
}
%le pongo un nuevo label
\rput(1.5,5){\large \textbf{M}}

%
%tapo el label que ten\'{\i}a
%
\pscustom[linestyle=none,fillstyle=solid,fillcolor=white]{
\psline(5,1.2)(6,1.2)
\psline(6,1.2)(6,1.8)
\psline(6,1.8)(5,1.8)
\psline(5,1.8)(5,1.2)
}
%le pongo un nuevo label
\rput(5.5,1.5){\large \textbf{h/J}}
\rput(3.7,5.2){\includegraphics[width=0.17\textwidth]{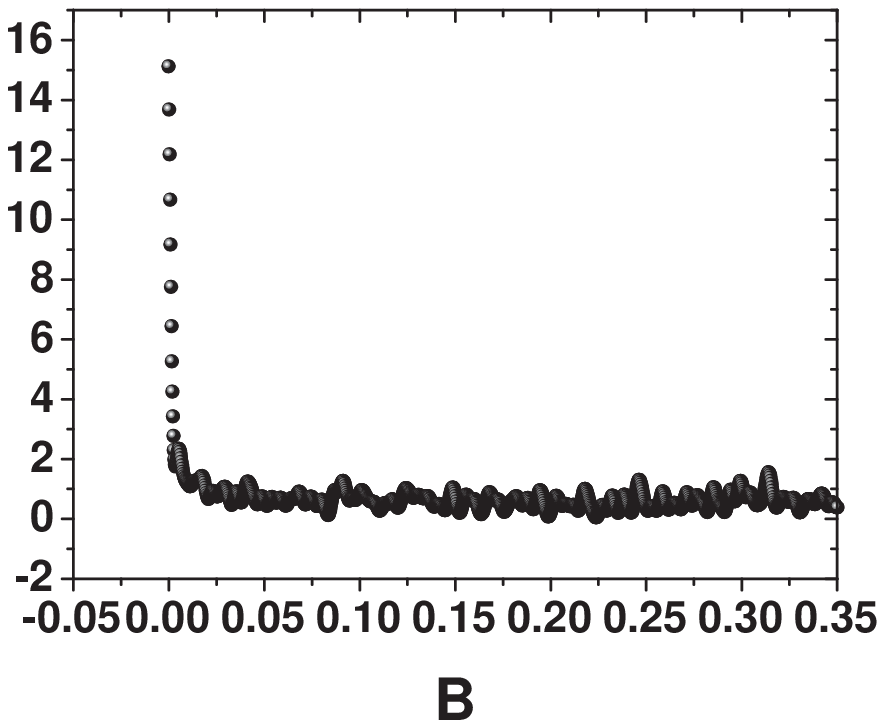}}
\rput(6,3){$J'/J=0.6$}
\rput(6,2.5){ $\sigma/J=0.5$}
\pscustom[linestyle=none,fillstyle=solid,fillcolor=white]{
\psline(3.9,4.3)(4.7,4.3)
\psline(4.7,4.3)(4.7,4.5)
\psline(4.7,4.5)(3.9,4.5)
\psline(3.9,4.5)(3.9,4.3)
}
\pscustom[linestyle=none,fillstyle=solid,fillcolor=white]{
\psline(2.81,4.3)(3.6,4.3)
\psline(3.6,4.3)(3.6,4.5)
\psline(3.6,4.5)(2.81,4.5)
\psline(2.81,4.5)(2.81,4.3)
}
\pscustom[linestyle=none,fillstyle=solid,fillcolor=white]{
\psline(3.7,4.35)(4.1,4.35)
\psline(4.1,4.35)(4.1,3.9)
\psline(4.1,3.9)(3.7,3.9)
\psline(3.7,3.9)(3.7,4.35)
}

\pscustom[linestyle=none,fillstyle=solid,fillcolor=white]{
\psline(2.5,6)(2.6,6)
\psline(2.6,6)(2.6,4.8)
\psline(2.6,4.8)(2.5,4.8)
\psline(2.5,4.8)(2.5,6)
}
\pscustom[linestyle=none,fillstyle=solid,fillcolor=white]{
\psline(2.5,4.6)(2.6,4.6)
\psline(2.6,4.6)(2.6,4.5)
\psline(2.6,4.5)(2.5,4.5)
\psline(2.5,4.5)(2.5,4.6)
}
\rput(2.55,5.7){\tiny $\chi$}
\rput(3.8,4.2){\tiny \textbf{h/J}}
\end{pspicture}

\caption{\label{devnum}  Magnetization curve for a disordered chain with $L=102$ sites and $J'/J=0.6$,
averaged over 150 samples. The magnetization drops quickly to zero rendering a singular magnetic
susceptibility at zero field, as shown in the inset.}
\end{figure}

\begin{figure}[t]
\begin{pspicture}*(1,0.5)(11,7.7)
%\psgrid
\centering
\rput(4.6,5.){\includegraphics[width=0.72\textwidth]{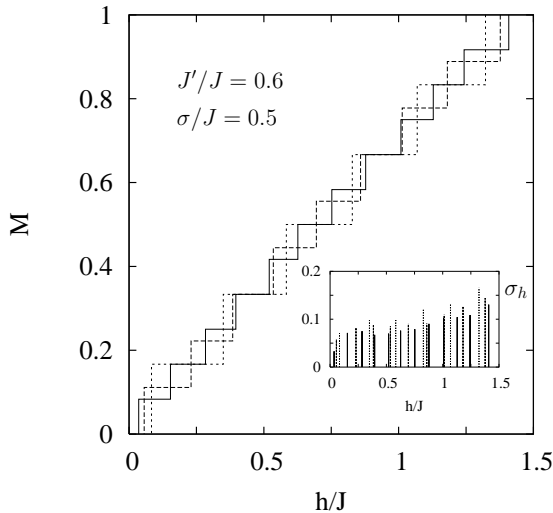}}
\end{pspicture}\\
\caption{\label{diag_desorden} Exact magnetization curves of disordered chains
averaged over 100 samples for $L =$ 24, 18 and 12 spins (solid, dashed and dotted
lines respectively;
notice that averages were taken for the critical fields, not for the allowed magnetization values,
so the curve is not smoothed).
$J, J'$ and $\sigma$ are taken as in Fig.\,\ref{devnum}
The inset denotes the standard deviations $\sigma_h$ of the
corresponding critical fields.  }
\end{figure}

We focused particular attention on low magnetic fields, so as to
detect possible singularities near $h = 0$ as those observed in
disordered NN chains \cite{Cabra}. In Fig.\ \ref{devnum} we
display the magnetization curve for 102 spins  with $J'/J=0.6$ and
$\sigma/J = 0.5$. Notice that  the magnetization plateau observed
in the pure case ($\sigma = 0$) is now totally suppressed. In
fact, all studied values of $\sigma$ suggest that the plateau of
the pure system is unstable under disorder. This observation was
also corroborated in smaller systems after diagonalizing them exactly
over 100 disorder realizations. In Fig.\ \ref{diag_desorden} we show for
comparison the magnetization curves obtained for the same values
of $J,\, J'$ and $\sigma$. On the contrary, the magnetization
remains finite and drops quickly to zero at zero field. Actually,
the magnetic susceptibility $\chi=\frac{\partial M}{\partial h}$
exhibits a divergence at $h=0\,$, as shown in the inset of Fig.\
\ref{devnum}.

A thorough exploration of mean values of NN and NNN couplings and
disorder strength shows that the low field magnetization curve
shows a behavior compatible with a power law $M(h)\sim h^{\alpha}$
in most of the parameter space, except on a small region $J'/J
\alt 10^{-4}$ where $M$ decreases in a logarithmic form $M\sim
\frac{1}{(\log(h^{2}))^{2}}$. In the power law region an exponent
$\alpha <1$ is obtained for disorder strength $\sigma/J \agt 0.3$,
corresponding to a singularity in the zero field magnetic
susceptibility. For $\sigma/J \alt 0.3$ the magnetization
decreases with an exponent $\alpha$ generally larger than one.

%

%_________________ Nueva Figura ________________
%____________________________________________

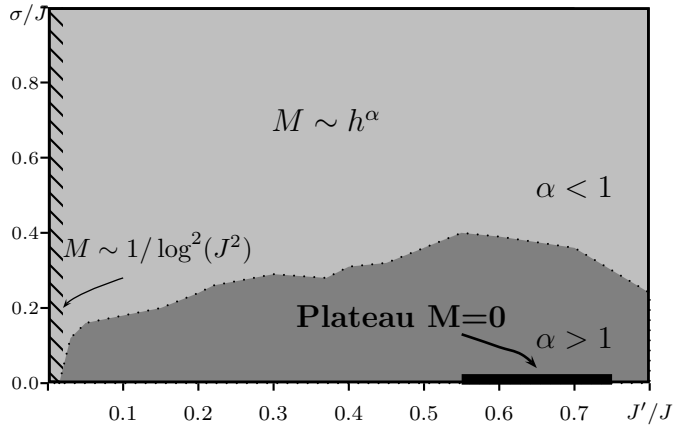
\begin{figure}[t]
\begin{pspicture}*(0.5,0)(9.5,6.3)
%\psgrid
% caption
\psframe[linewidth=1pt,fillstyle=solid,fillcolor=lightgray](1,1)(8,6)
%[linewidth=1pt,fillstyle=solid,fillcolor=gray]
\psframe[linewidth=1pt,fillstyle=solid,fillcolor=lightgray](1,1)(9,6)
%\psframe[linewidth=1pt,fillstyle=solid,fillcolor=gray](1,1)(10,8)
%\pscustom[linewidth=2pt,fillstyle=solid,fillcolor=gray]{
%\pscurve[linewidth=3pt](1.1,6)(2,2)(6,1.2)
% labels de los ejes
\rput(10.9,0.4){$J'/J$}
\rput(0.75,5.9){\textbf{{\footnotesize $\sigma/J$}}}
% rayitas del eje x
\psline{-}(5,0.9)(5,1.0)
\psline{-}(1,0.9)(1,1.0)
\psline{-}(2,0.9)(2,1.0)
\psline{-}(3,0.9)(3,1.0)
\psline{-}(4,0.9)(4,1.0)
\psline{-}(6,0.9)(6,1.0)
\psline{-}(7,0.9)(7,1.0)
\psline{-}(8,0.9)(8,1.0)
\psline{-}(9,0.9)(9,1.0)
\psline{-}(10,0.9)(10,1.0)
% rayitas del eje y
\psline{-}(0.9,1)(1.0,1)
\psline{-}(0.9,2)(1.0,2)
\psline{-}(0.9,3)(1.0,3)
\psline{-}(0.9,4)(1.0,4)
\psline{-}(0.9,5)(1.0,5)
\psline{-}(0.9,6)(1.0,6)
%escala x
\footnotesize
\rput(2,0.6){$0.1$}
\rput(3,0.6){$0.2$}
\rput(4,0.6){$0.3$}
\rput(5,0.6){$0.4$}
\rput(6,0.6){$0.5$}
\rput(7,0.6){$0.6$}
\rput(8,0.6){$0.7$}
\rput(9,0.6){$J'/J$}
\rput(10,0.6){$0.9$}
%
%escala y
\scriptsize
\rput(0.7,1){\textbf{$0.0$}}
\rput(0.7,2){\textbf{$0.2$}}
\rput(0.7,3){\textbf{$0.4$}}
\rput(0.7,4){\textbf{$0.6$}}
\rput(0.7,5){\textbf{$0.8$}}
\normalsize
%DIBUJANDO EL DIAGRAMA DE FASES
%
%regi\'{o}n donde no diverge
\pscustom[linestyle=dotted,fillstyle=solid,fillcolor=gray]{
\psline[linewidth=1pt]{-}(1.15,1)(1.3,1.6)(1.5,1.8)(2.5,2)(3,2.2)(3.2,2.3)(4,2.45)(4.7,2.4)(5,2.55)(5.5,2.6)(6,2.8)(6.5,3)(7,2.95)(8,2.8)(9,2.2)
\psline(9,1)(1,1)}
\rput(8,1.6){\large $\alpha>1$}
\rput(8,3.6){\large $\alpha<1$}
%\pscustom[linewidth=1pt,fillstyle=solid,fillcolor=gray]{
%\pscurve(0,2)(1,1.25)(2,1.5)(4,3)
%\pscurve(4,1)(3,0.5)(1,0)(0,5)}
%
%plateau M=0
\psline[linewidth=4pt]{-}(6.5,1.05)(8.5,1.05)
\pscurve[linewidth=1pt]{<-}(7.5,1.2)(7.3,1.38)(7,1.48)(6.8,1.55)(6.5,1.65)
\rput(5.7,1.9){\large \textbf{Plateau M=0}}
%
%regi\'{o}n universal
%\pscustom[linestyle=none,fillstyle=vlines,fillcolor=darkgray]{
%\psline(5,5)(10,5)
%\psline(10,3.5)(5,3.5)
%\psline(5,3.5)(5,5)
%}
%\rput(7.5,4.3){ \Large \textbf{Regi\'{o}n Universal ?}}
\pscurve[linewidth=0.5pt]{->}(2,2.4)(1.3,2.1)(1.2,2)
%
%fase RS
\pscustom[linestyle=none,fillstyle=vlines]{
\psline[linecolor=black](1.02,1)(1.02,6)
\psline[linecolor=black](1.02,6)(1.2,6)
\psline[linecolor=black](1.2,6)(1.2,1.02)
\psline[linecolor=black](1.2,1.02)(1.02,1.02)
}
\rput(4.7,4.5){\large  \textbf{$M \sim h^{\alpha}$}}
\rput(2.4,2.8){  \textbf{$M \sim 1/\log^{2}(J^{2})$}}
%\pscurve[linecolor=white]{->}(5,5.8)(4,5.5)(3,5)
%cosas que sobran
%\cput*{0}(2.5,3.5){$RS$}
%\cput*{45}(4.0,5.0){$RD$}
%\cput*{45}(1.5,1.5){$RD$}
%\pscircle(2.5,3.5){0.4}
%\psline{<-}(1.6,2.6)(2.2,3.2)
\psframe[linewidth=1pt,fillstyle=none,fillcolor=lightgray](1,1)(9,6)
\end{pspicture}
\caption{\label{diagrama} Schematic diagram of the low field
results obtained for disordered NNN chains. In a tiny (dashed)
region the magnetization is logarithmic like. In contrast, in most
of parameter space the magnetization follows a power law, with
larger disorder leading to singular susceptibility (light gray
region) and small disorder leading to smooth magnetization (gray
region). A plateau at $M=0$ is present only in the absence of
disorder.}
\end{figure}

In Fig.\ \ref{diagrama} we show a schematic diagram of our SCMF
numerical results. The dashed zone  denotes a logarithmic
behavior, the light gray region represents a power law behavior
decrease of $M$ with exponent $\alpha < 1$, whereas in the gray zone
this exponent is larger than one. For the ordered case $\sigma=0$
we show a bold line representing the magnetic plateau at $M=0$
within $0.55 <  J'/J < 0.75$\,.

We can finally compare the low field behavior of disordered
antiferromagnetic NNN chains with those found in disordered NN AF-F
chains (Section II). We can stress that  both systems have two
phases: a dominant one characterized by a power law magnetization
behavior and a small region of the parameter space where that
behavior is close to a logarithmic type. Moreover, in the power
law regime both systems can develop singular or smooth zero field
magnetic susceptibilities.

%
%
%
%
%
%%%%%%%%%%%%%%%%%%%%%%%%%%%%%%%%%%%%%%%%%%%%%%%%%%%%%%%%%%%%%%%%%%%%%%%%%%%%%%%%%%%%%%%%%
%%%%%%%%%%%%% DIAGRAMA DE RESULTADOS %%%%%%%%%%%%%%%%%%%%%%%%%%%%%%%%%%%%%%%%%%%%%%%%%%%%

\section{Summary and conclusions}
\label{summary}

In the first part of this work we have studied analytically $XX$
spin-1/2 chains with random NN interactions both antiferromagnetic
and ferromagnetic, by suitably extending the analysis based on a
random walk problem presented in Ref.\ [\onlinecite{egg}].
We have distinguished
three regions in the parameter space with different low magnetic field
behavior: singular power law $ M \propto h^{\alpha}$ with
$\alpha < 1$, smooth power law with $\alpha > 1$ and a logarithmic
dependence $\propto \frac{1}{(\log(h^{2}))^{2}}$. The second part
discusses an alternative approach to random antiferromagnetic
spin-1/2 chains which allows for NN and NNN interactions, using a
numerical self consistent mean field method adapted from
Ref.\ [\onlinecite{rossini}]. As for the NN chains analyzed before, we
have found three phases in these systems, the dominant one being  a
power law dependence of the magnetization with the low external field
in most of the parameter space. Also, a logarithmic dependence of
the magnetization was found just within a small region of
parameters.

Our results show that the two systems have similar low energy
properties. A similar analogy has been found in the study of the
same two systems for the Heisenberg ($SU(2)$) version
\cite{Hoyos,Yusuf}. In both of them there are three phases and
most of the parameter space is dominated by a power law dependence
of $M(h)$. Both systems also show a small region where the
magnetization displays the same kind of logarithmic singularity.
In both systems the power law region follows a dynamical exponent
$\alpha > 1$ for weak disorder, thus yielding a well behaved
magnetic susceptibility. On the contrary, strong disorder yields
exponents smaller than one and consequently the susceptibility
becomes singular at  $h=0$. This later behavior has been also
found in dimerized spin-1/2 chains with disordered NN interactions
\cite{Hyman}.

\acknowledgments We thank A.\ Honecker for helpful discussions.
This work was partially supported by ECOS-Sud Argentina-France
collaboration (Grant A04E03), PICS CNRS-CONICET (Grant 18294),
PICT ANCYPT (Grant 20350), and PIP CONICET (Grant 5037).

%%%%%%%%%%%%%%%%%%%%%%%%%%%%%%%%%%%%%%%%%%%%%%%%%%%%%%%%%%%%%%%%%%%%%%%%%%%%%%%%%%%%%%%%%%%
%%%%%%%%%%%%%%%%%%%%%%%%%%%%%%%%%%%%%%%%%%%%%%%%%%%%%%%%%%%%%%%%%%%%%%%%%%%%%%%%%%%%%%%%%%%
%
%       BIBLIOGRAF\'{I}A (?` no autorizada ?)
%
%%%%%%%%%%%%%%%%%%%%%%%%%%%%%%%%%%%%%%%%%%%%%%%%%%%%%%%%%%%%%%%%%%%%%%%%%%%%%%%%%%%%%%%%%%%%
%%%%%%%%%%%%%%%%%%%%%%%%%%%%%%%%%%%%%%%%%%%%%%%%%%%%%%%%%%%%%%%%%%%%%%%%%%%%%%%%%%%%%%%%%%%%
%

\end{document}